\documentclass[5p,twocolumn]{elsarticle}
\usepackage{graphicx,latexsym}
\usepackage{dcolumn}
\usepackage{amssymb,amsmath,bm}
\usepackage{subfigure}
\usepackage{braket}
\usepackage{siunitx}
\usepackage{array}
\usepackage{float}
\usepackage[nopar]{lipsum}
\usepackage{caption}
\usepackage{multirow}
\usepackage{bigstrut}
\usepackage[super]{nth}

\biboptions{sort&compress}

\newcommand{\angstrom}{\text{\normalfont\AA}}
\usepackage{hyperref}
\hypersetup{
    pdfnewwindow=true,       
    colorlinks=true,         
    linkcolor=blue,          
    citecolor=blue,          
    filecolor=magenta,       
    urlcolor=black           
}

\usepackage[normalem]{ulem}

\def\sec#1{Sec.\ \ref{#1}}

\def\fig#1{Fig.\ \ref{#1}}

\journal{}

\begin{document}

\begin{frontmatter}


\title{Electronic and Optical properties of Metallic Nitride: A comparative study between the MN (M=Al, Ga, In, Tl) monolayers}

\author[a1,a2]{Nzar Rauf Abdullah}
\ead{nzar.r.abdullah@gmail.com}
\address[a1]{Division of Computational Nanoscience, Physics Department, College of Science, 
             University of Sulaimani, Sulaimani 46001, Kurdistan Region, Iraq}
\address[a2]{Computer Engineering Department, College of Engineering, Komar University of Science and Technology, Sulaimani 46001, Kurdistan Region, Iraq}

\author[a3]{Botan Jawdat Abdullah}
\address[a3]{Physics Department, College of Science- Salahaddin University-Erbil, Erbil 44001, Kurdistan Region, Iraq}

\author[a4]{Vidar Gudmundsson}
\ead{vidar@hi.is}
\address[a4]{Science Institute, University of Iceland, Dunhaga 3, IS-107 Reykjavik, Iceland}


\begin{abstract}
The electronic and the optical properties of metallic nitride (MN) monolayers are studied using a DFT formalism.
In most of these monolayers, the electron density of the metallic atoms is much higher than that of the nitride atoms, and ionic, covalent, and metallic bonds are found in M-N bonds, resulting in fascinating electronic and optical properties.
The optical band gap is varied from almost $0.0$ to $3.0$~eV for the MN monolayers depending on 
the bond type between the metallic and the nitride atoms, as well as the contribution of the type of orbitals around the Fermi energy. The optical properties such as the dielectric function, the excitation spectra, the refractive index, the reflectivity, and the optical conductivity of MN monolayers are calculated. The excitation energy and static dielectric constant are found to be inversely proportional to the band gap at low photon energy. The MN monolayers with a large band gap have good visible light functionality, while the MN monolayers with a lower band gap are found to be active in the infrared region. Furthermore, it is shown that the optical properties of MN monolayers show a strong anisotropy with respect to the polarization of the incoming light. Consequently, our results for the optical properties of MN monolayers show that they could be beneficial in optoelectronic device applications.
\end{abstract}

\begin{keyword}
Metallic Nitride monolayers \sep DFT \sep Electronic structure \sep  Optical properties
\end{keyword}

\end{frontmatter}

\section{Introduction} 

Two-dimensional (2D) nanomaterials have emerged as a class of materials with very distinct structure and physical properties. The research of 2D materials has lately made significant progress, with applications in next-generation nanoscale technology for the future \cite{Novoselov2005, Li2021, Wang2019, RASHID2019102625, Tan2020}. Scientists have been working on graphene, the first 2D material, continuously, both theoretically and experimentally, since its successful production \cite{doi:10.1126/science.1102896, ABDULLAH2020126350}. The gapless nature of graphene, on the other hand, has proven to be a significant impediment to its application. As a result, there is a pressing need to develop novel 2D materials that have both outstanding graphene-like properties and a moderate band gap \cite{ABDULLAH2020100740}.

Other graphene-related compounds are reportedly currently being investigated in order to extend the range of device applications \cite{Balendhran2015, Xu2013, doi:10.1021/acs.chemrev.6b00558, Prete2020}. Among them, group III-V nitrides have long been regarded as a possible semiconductor materials. Boron Nitride (BN), Aluminum Nitride (AlN), Gallium Nitride (GaN), Indium Nitride (InN) and Thallium Nitride (TlN) are examples of graphene-like 2D III-nitrides. Because Al, Ga, In, and Tl are metals in the group III nitrides, and boron is a metalloid, we choose to investigate metallic nitride (MN) monolayer by assuming (M=Al, In, Ga, and Tl) in this study. There are also more recent experimental \cite{doi:10.1063/1.3041639, pub.1014753295} and theoretical \cite{PhysRevB.79.115442, C2CP40081B, C2NR32366D} studies on BN monolayers. 

The fabrication of 2D materials, particularly AlN, InN, GaN, and TlN, has attracted researchers' attention. The AlN monolayer was recently synthesized using a standard chemical vapour deposition technique \cite{ZHANG2007317}. Plasma assisted molecular beam epitaxy on Ag(111) single crystals was used to produce the AlN monolayer epitaxially \cite{doi:10.1063/1.4851239}, and the AlN monolayer was produced on the (111) Silicon substrate via the RHEED technique in ammonia molecular beam epitaxy \cite{Mansurov2015}. The GaN monolayer has been experimentally fabricated using a hybrid carbon nanotube and graphene structure by metal-organic chemical vapor deposition \cite{Seo2015}, and by a graphene encapsulation method \cite{AlBalushi2016}.

Moreover, several InN nanostructures, such as InN nanotubes, nanowires \cite{doi:10.1063/1.2712801, doi:10.1021/nl4030819}, and monolayer InN quantum wells \cite{doi:10.1063/1.4967928}, have been experimentally synthesized. Although TlN has not been prepared, a theoretical investigation using DFT indicated that it has the same structure as other metal nitrides \cite{SHI2010203, Elahi2016, Wei2010}, implying that it might be synthesized.

Some of these compounds have been modeled theoretically, particularly the electronic and the optical characteristics for a certain range of parameters. For example, the electrical and the optical properties of an AlN nanosheet have been explored using density functional theory (DFT). The dielectric function, the absorption coefficient, the optical conductivity and the extinction index of a AlN nanosheet have been evaluated and the results show that it has semiconductor characteristics \cite{VALEDBAGI2013153}.
Based on the first principle many-body Green's function and Bethe–Salpeter equation approach, the electronic structure and the optical characteristics of an AlN monolayer have been computed. The enhanced excitonic effects in the AlN monolayer can be utilized in nano-optoelectronic devices, according to the findings \cite{VAHEDIFAKHRABAD201538}.
Chemical functionalization increases the band gaps of GaN monolayers, according to first-principles calculations and it allows rapid modification of electrical and optical properties. The increase in visible-light absorption is confirmed by the   absorption spectra of the half-hydrogenated GaN monolayer, except for the half-fluorinated monolayer \cite{Meng2016, C8CP05529G}.

The electrical and optical properties of pure and chemically functionalized InN monolayers with Cl and F atoms have been investigated with first-principle computations. The energy gap of these monolayers is direct and enhanced. The dielectric function of the chemically functionalized InN monolayers leads to a high absorption coefficient in the visible light region \cite{D0RA01025A}. In addition, the electrical and optical characteristics of a TlN monolayer have been calculated for limited parameters \cite{FERREIRADASILVA2005151}, and DFT was used to study the dielectric function, the optical conductivity, the extinction index, and the energy loss function of TlN nanosheets for two directions of linear polarization. The optical conductivity of a TlN monolayer with light polarized parallel or perpendicular to the sheet suggests that it has semiconducting properties \cite{Elahi2016}.

In this work, the electrical and the optical properties of MN monolayers are studied. We compute all important factors related to the electrical and the optical properties in this investigation such as the electron density, the band structure, the density of states, and the optical properties. The optical properties of MN monolayer compounds, according to the study, are important for possible optoelectronic applications.

The computational methodologies and model structure are briefly discussed in \sec{Sec:Computational}. The major achieved outcomes are examined in \sec{Sec:Results}. The conclusion of the results is reported in \sec{Sec:Conclusion}

\section{Methodology and computational tools}\label{Sec:Computational}

We consider 2D dimensional metallic nitrite monolayers confined in the $xy$-plane. 
The DFT calculations are performed using the Perdew-Burke-Ernzerhof (PBE) form of the generalized gradient approximation (GGA) \cite{ABDULLAH2021110095} for the exchange correlation functional implemented in the Quantum Espresso (QE) simulation package \cite{Giannozzi_2009, giannozzi2017advanced}. 

For the structure calculation a cutoff for the plane-waves is set for the kinetic energy at $1088.5$~eV \cite{ABDULLAH2020126807}.
A vacuum layer is also assumed to be $20 \, \angstrom$ in the $z$-direction to avoid interaction that could take place from the adjacent metallic nitrite layer along this direction. 
The atomic position and cell parameters are considered fully optimized when all forces on the atoms are less than $10^{-5}$ eV/$\angstrom$ on the $18 \times 18 \times 1$ Monkhorst-Pack grid.

The same aforementioned parameters are used for self consistent field (SCF) and density of state (DOS) calculations, except that the momentum space is considered on a $100 \times 100 \times 1$ mesh for DOS calculation. The partial density of states (PDOS) is broadened by Tetrahedra smearing of $0.01$ eV. The optical properties of the monolayers are calculated using QE with an optical broadening of $0.1$~eV \cite{ABDULLAH2020103282}.

\section{Results}\label{Sec:Results}
\begin{figure}[htb]
	\centering
	\includegraphics[width=0.22\textwidth]{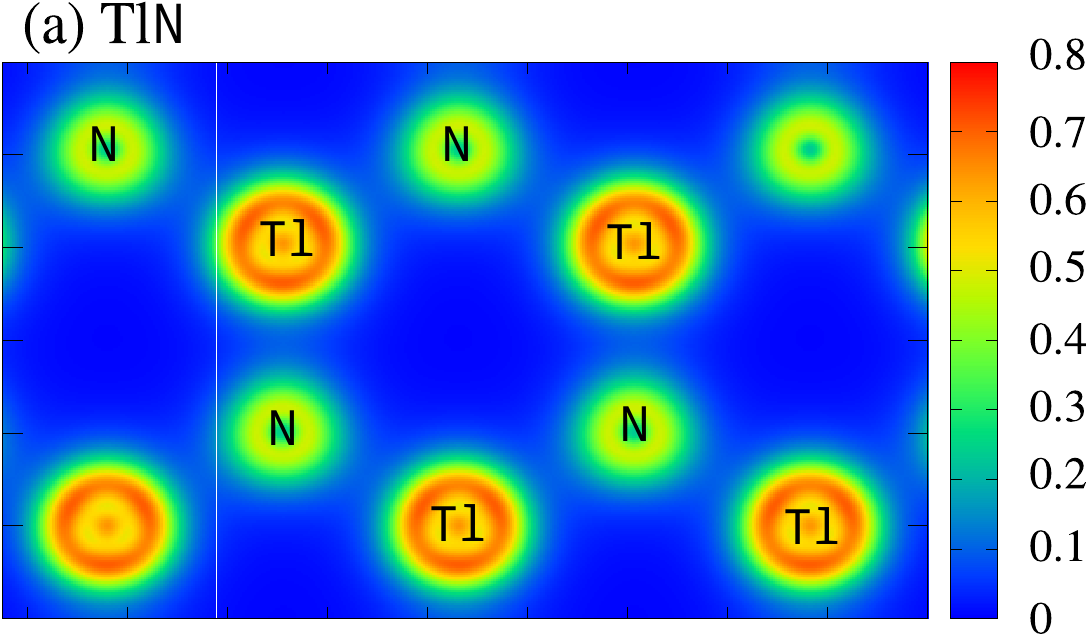}
	\includegraphics[width=0.22\textwidth]{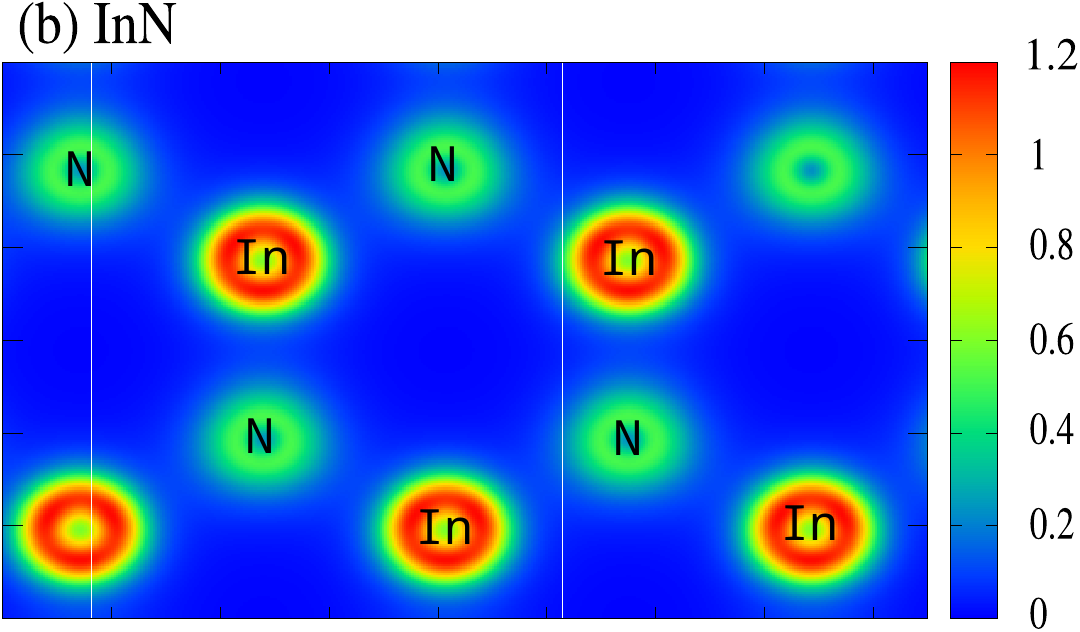}\\
	\includegraphics[width=0.22\textwidth]{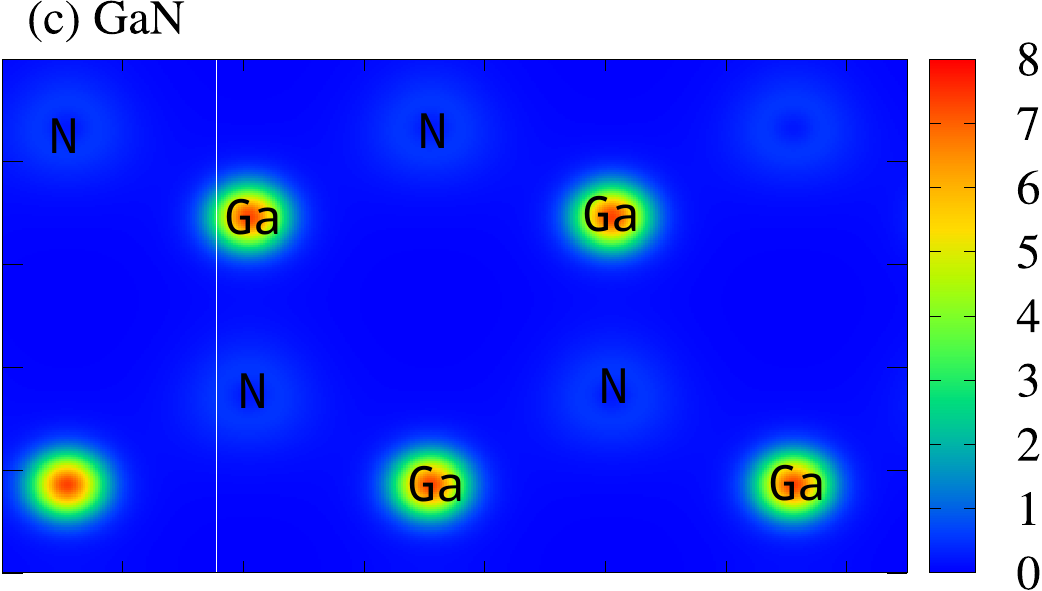}
	\includegraphics[width=0.22\textwidth]{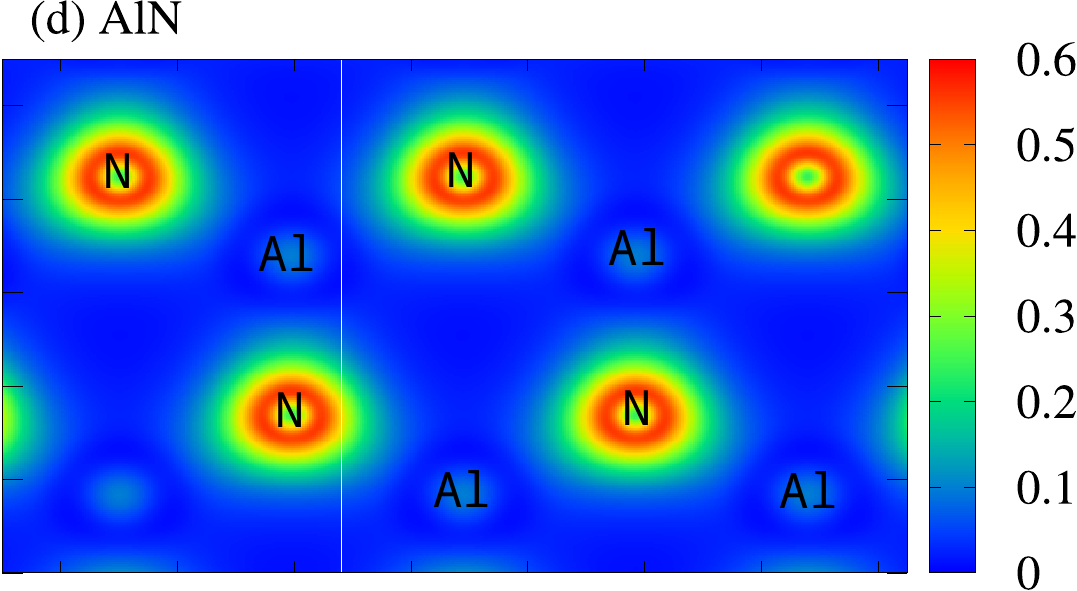}\\
	\caption{Electron density distribution of TlN (a), InN (b), GaN (c), and AlN (d).}
	\label{fig01}
\end{figure}

\subsection{Atomic structure and electron density}
We consider four $2\times2\times1$ supercells of metallic nitride (MN) monolayers, where M is assumed to be Tl, In, Ga, and Al atoms, giving rise to thallium nitrade (TlN), indium nitride (InN), gallium nitride (GaN), and aluminum nitride (AlN) monolayers.
In a hexagon of these monolayers, three nitrogen atoms and three M atoms exist, forming
a coplanar hexagonal ring with periodic boundary conditions in the plane.
The electron density distribution is presented in \fig{fig01} for fully relaxed structures of TlN (a), InN (a), GaN (c), and AlN (d) monolayers. In most of these monolayers, the electron density of the metallic atoms is much higher than for the nitrogen atoms. The electronegativity of an N atom is $3.04$, which is higher that that of the Tl, In, Ga, and Al atoms with the electronegativity values of $1.62$, $1.78$, $1.81$, and $1.61$, respectively.
\begin{figure}[htb]
	\centering
	\includegraphics[width=0.45\textwidth]{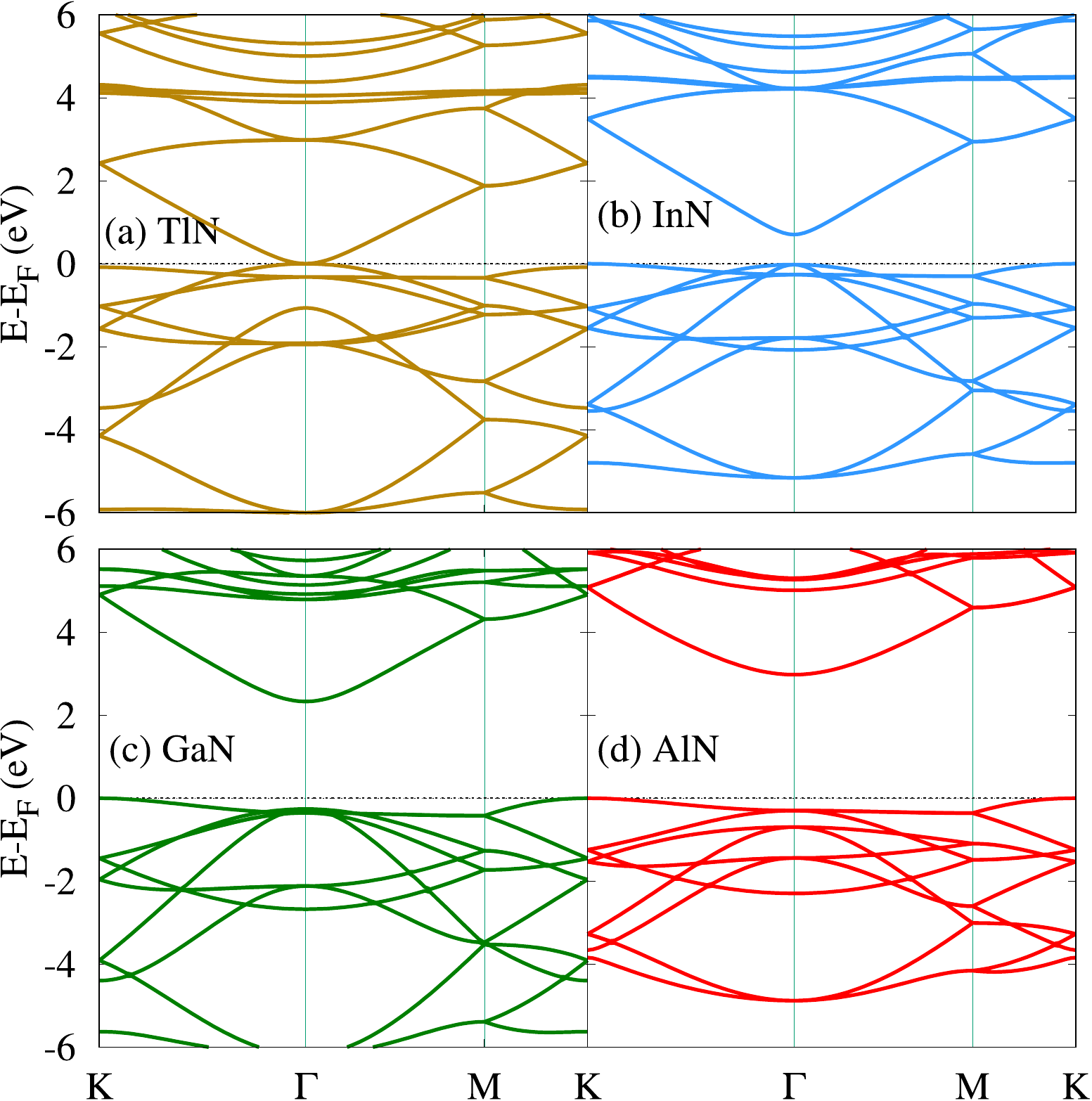}
	\caption{Band structure for optimized TlN (a), InN (b), GaN (c), and AlN (d) monolayers. 
		The Fermi energy is set to zero.}
	\label{fig02}
\end{figure}

The electron density distribution of TlN and InN monolayers indicates partly ionic bonds of the Tl and the In atoms with the N atoms, which is due to electron transfer from the Tl and the In atoms to the N atoms. The electron transfer occurs because the electronegativity of the N atoms is larger than that of the Tl and In atoms. In addition, the Tl-N and the In-N bonds are partly covalent bonds and partly metallic bonds due to the hybridization of the Tl-$p$, the In-$p$, and the N-$p$-orbitals, and the existing density of states around the Fermi energy, respectively (shown later). So, the nature of the Tl-N, and the In-N bonds shows mixture of ionic, covalent, and metallic character in the TlN and the InN monolayers.

The electron density of the AlN mnolayer shows bonds with more ionic behavior due to the electron transfer from Al to N leading to wavefunctions localized at the N-atoms with no significant overlap of the two neighboring atoms, and a high difference in electronegativities of Al and N atoms.

In the GaN monolayer, a localization of the electronic wavefunctions around Ga atoms is seen but with a smaller degree of distortion together with a smaller overlap of the wavefunctions centered at Ga atoms. As a results, the bonding in GaN is assumed to be partly ionic and partly covalent.

The most energetically favorable structures among these four monolayers can be determined by the bond lengths. The bond length is inversely proportional to the stability of a monolayer. 
The  Tl–N, In-N, Ga-N, and Al-N bond lengths are $2.14$, $2.05$, $1.83$, $ 1.799 \, \angstrom$ agreeing well with other DFT calculations \cite{VALEDBAGI2013153, D0RA01025A, ELAHI20169367}.
Consequently, the monolayers can be arranged from the high to the low energetically stable structure as follows: AlN, GaN, InN, and TlN monolayers. The bond length influences the lattice constant in the sense that the longer the bond length, the longer is the lattice constant obtained. 
The lattice constant of TlN, InN, GaN, and AlN is $3.72$, $3.55$, $3.18$, and $3.11$~$\angstrom$, respectively, in which the lattice constant of the TlN monolayer is the largest. This is expected as the atomic radius of the Tl atom is $1.56$~$\angstrom$, which is larger than the atomic radii of the Ga and the Al atoms.
\begin{figure}[htb]
	\centering
	\includegraphics[width=0.45\textwidth]{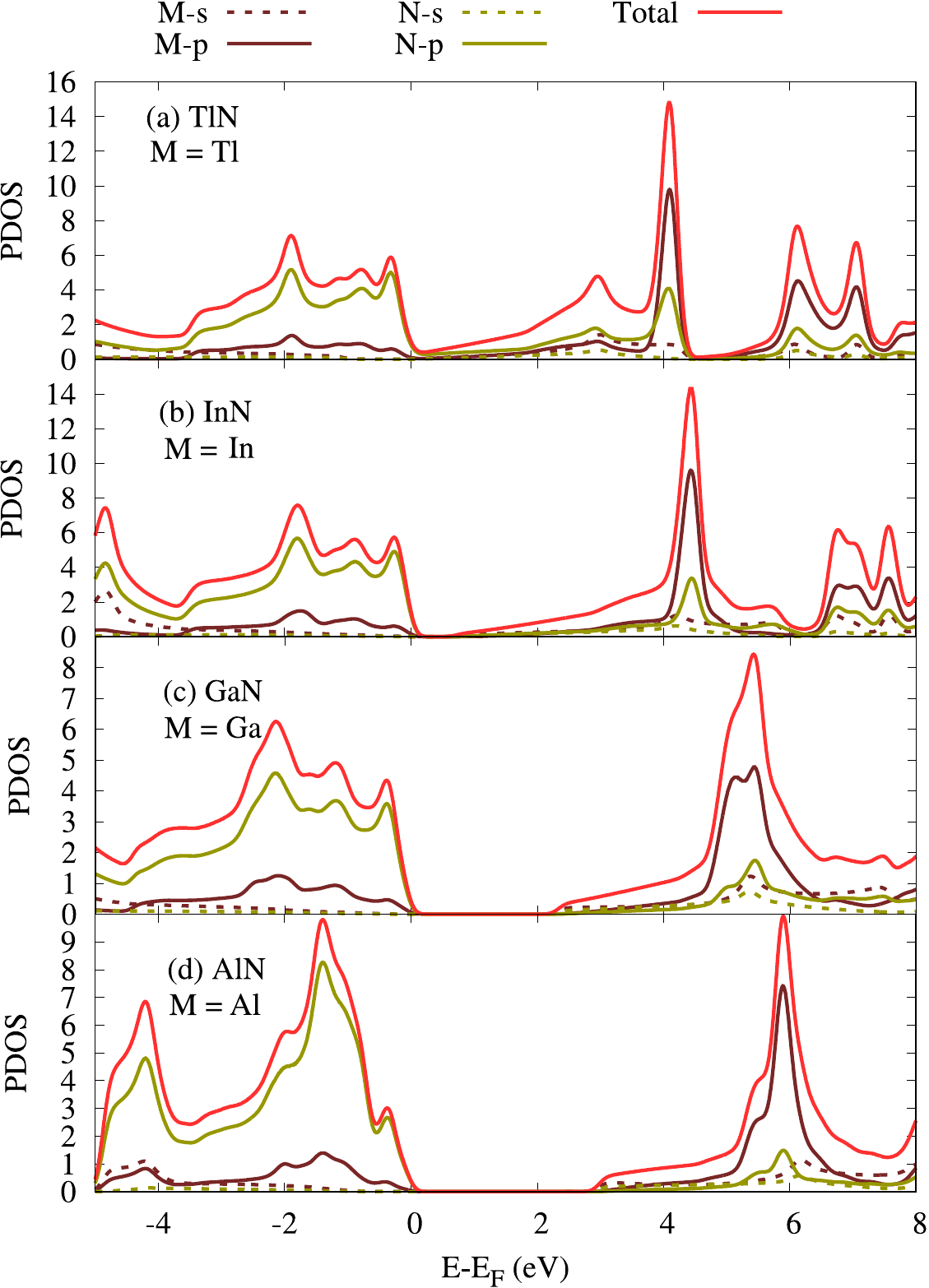}
	\caption{Partial density of states (PDOS) of  TlN (a), InN (b), GaN (c), and AlN (d). The PDOS of s- (dashed lines), the p-orbital (solid golden and blood lines) of all atoms (N and M = Tl, In, Ga, Al) are plotted with the total DOS of the monolayers (solid red line). The Fermi energy is set to zero.}
	\label{fig03}
\end{figure}

The formation energy is another approach to determine the monolayers' stability. 
\lipsum[0]
\begin{figure*}[htb]
	\centering
	\includegraphics[width=0.9\textwidth]{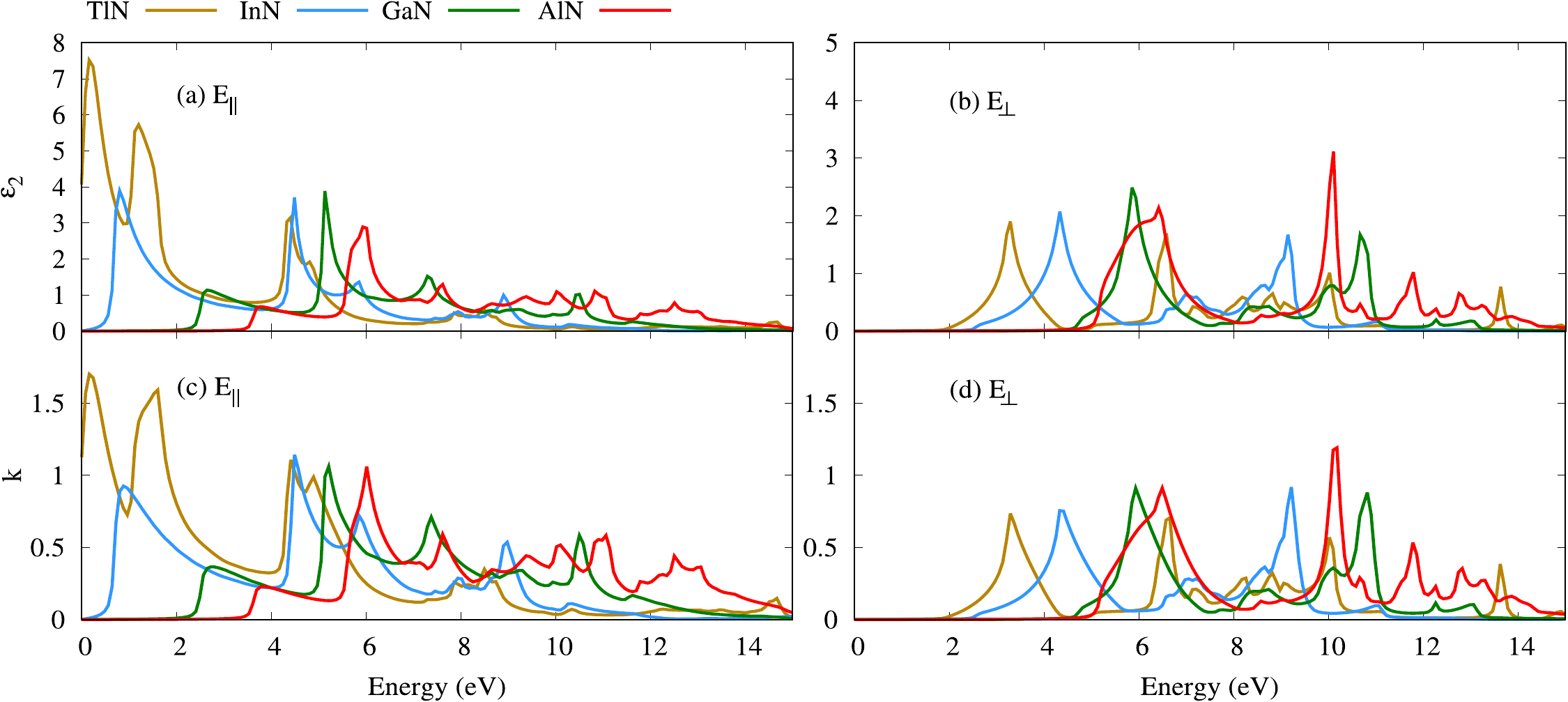}
	\caption{Imaginary part of the dielectric function, $\varepsilon_2$ (a) and (b), and excitation spectra, $k$ (c) and (d), for a parallel $E_{\rm \parallel}$ (left panel) and perpendicular $E_{\rm \perp}$ (right panel) polarizations of the electric field, respectively.}
	\label{fig04}
\end{figure*}
Formation energy is the energy required for producing an
atomic configuration of a monolayer which demonstrates it's energetic stability \cite{ABDULLAH2020114556, ABDULLAH2021114644}.
The DFT calculations show that the formation energy is -4.402, -6.052, -7.594, and 
-10.168~eV for TlN, InN, GaN, and AlN monolayers, respectively.

It has been shown that the smaller the formation energy, the more energetically stable the monolayer should be. Consequently, these monolayers can be arranged from the most stable to the less stable structure as follows: AlN, GaN, InN, and TlN \cite{Liu2019}.

\subsection{Electronic band structure and density of state}
The electronic band structures of the TlN (a), the InN (b), the GaN (c), and the AlN (d) monolayers are presented in \fig{fig02}, where all the bands are shifted so that the Fermi energy is set to be zero.
Only the bands within the range of $6$~eV around the Fermi energy are shown around the high symmetrical k-points of K, $\Gamma$, and M \cite{ABDULLAH2021106073}.
We have found a direct band gap of TlN with energy $2.1$ meV at the $\Gamma$ point, and indirect band gaps in InN, GaN, and AlN with energies $0.702$, $2.33$, and $2.97$~eV along $\Gamma$-K path, respectively. This indicates that TlN has a semi-metal property, but the InN, GaN, and AlN monolayers behave as semiconductors.

The band gaps are underestimated as the GGA-PBE has been used. In order to get more accurate band gaps of these monolayers, one may use Heyd–Scuseria–Ernzerhof (HSE06) hybrid functionals \cite{doi:10.1063/1.1564060} or use the GW approximations \cite{PhysRev.139.A796}. 
Using the HSE06 functional, the obtained band gaps of TlN, InN, GaN, and AlN are 
$0.13$, $1.72$, $3.27$, $4.03$~eV, respectively, which are in agree with other DFT results \cite{D0RA01025A, Meng2016, Liu2019}.

The results for the band gaps of these monolayers are in contrast to the band gaps of 
bulk wurtzite nitrides with the gaps being direct for AlN and GaN \cite{PhysRevB.50.4397}.
This difference in the band gap property of the monolayers and wurtzite nitrides can be related to
a shift in the position of the valence band maxima from $\Gamma$ to K, when the dimensionality of the system is reduced, while the property of the conduction band minima remain almost the same.

In order to get an insight into how the electrons of individual elements contribute to the band structures, we show the total density of state (TDOS) and partial density of states (PDOS) for TlN (a), InN (b), GaN (c), and AlN (d) in \fig{fig03}. 
We observe that the valence electrons come mostly from the hybridization of the $p$-orbitals of the M and the N atoms originating from the nitrogen atoms, and the conduction electrons come from the hybridization of the $p$-orbital of the M and the N atoms originating mainly from M atoms. Furthermore, electrons from the $s$-orbitals of the M and the N atoms in the conduction band region near the Fermi energy contribute to the band structures.   
We can more precisely confirm that the $p_z$ electrons are the majority of valence electrons, while the $s$ electrons have some population together with the $p$ electrons in the conduction band region.
The above analysis confirms that electrons from the $p_z$-orbitals of the N atoms in the valence band, 
and electrons of the $s$-orbitals of the M atoms in the conduction band region are critical in forming the band gap.

\subsection{Optical properties}
The optical properties of a structure demonstrate how it interacts with
the electromagnetic field of an incident light, and the
optical properties are directly related to the band structure of a material.
Consequently, we expect different optical properties for the different band structures and band gaps of the monolayers \cite{JOHN2017307}.
The real, $\varepsilon_1$, and the imaginary, $\varepsilon_2$, parts of
the dielectric function are calculated using the random phase approximation (RPA) \cite{Ren2012, ABDULLAH2020126578} implemented in the QE software.

\lipsum[0]
\begin{figure*}[htb]
	\centering
	\includegraphics[width=0.9\textwidth]{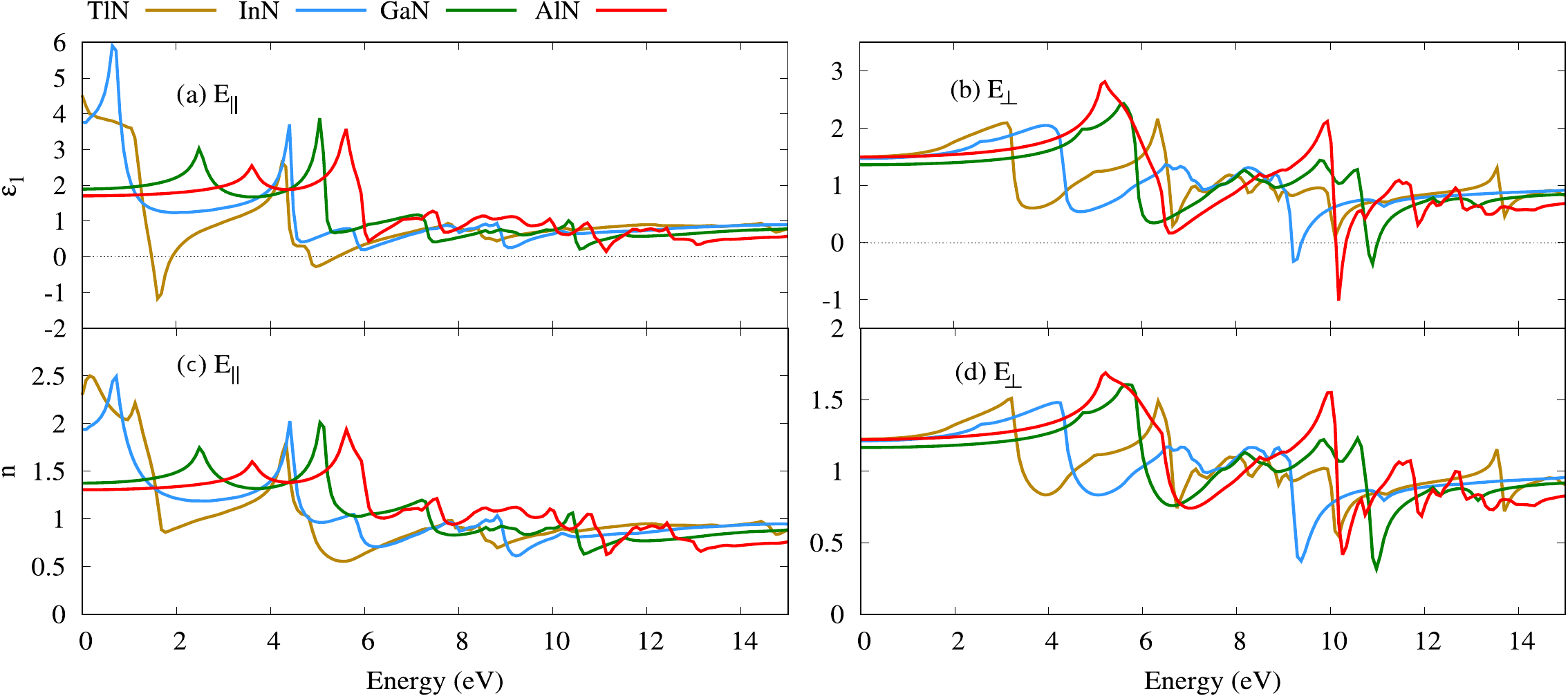}
	\caption{Real part of the dielectric function, $\varepsilon_1$ (a) and (b), and Refractive index, $n$ (c) and (d), for a parallel $E_{\rm \parallel}$ (left panel) and perpendicular $E_{\rm \perp}$ (right panel) polarizations of the electric field, respectively.}
	\label{fig05}
\end{figure*}

In \fig{fig04} are shown the $\varepsilon_2$, and the excitation spectra, $k$, for an MN monolayer with the electric field polarized parallel, E$_{\rm \parallel}$, (a, c), or perpendicular, E$_{\rm \perp}$, (b,d) to the plane of the structures. 
The simulated results show that while the optical gaps of
the GaN and AlN monolayers are in the energy of the
visible light region, the optical gaps of the TlN and InN monolayers are 
in the near-infrared region in the case of E$_{\rm \parallel}$.

We can extract the optical gap or the excitation energy of each monolayer from either the $\varepsilon_2$ or $k$ spectra. We notice that the optical gaps of the GaN and AlN monolayers are at $2.32$ and $3.0$~eV for the parallel incident light, respectively.
This implies that the GaN and AlN monolayers have a strong ability to absorb visible light. 
The optical gap of TlN and InN monolayers are at $0.001$, and $0.5$~eV, respectively, for E$_{\rm \parallel}$ indicating that the primary optical gaps of TlN and InN are in the infrared and near-infrared regions, respectively.

The optical gap for the case of E$_{\rm \perp}$ is at $1.85$ (visible), $2.39$ (visible), $4.58$ (UV), and $4.86$~eV (UV) for TlN, InN, GaN, and AlN monolayers, respectively. 
In contrast to E$_{\rm \parallel}$, the TlN and InN have a better absorbance ability of visible light in the case for E$_{\rm \perp}$. The results of the $\varepsilon_2$, and $k$ confirm the band structures and the band gaps of the MN monolayers.

The real part of dielectric function, $\varepsilon_1$ (a,b), and the refractive index, $n$ (c,d), of the MN monolayers are shown in \fig{fig05} for the electric field parallel (left panel), and perpendicular (right panel) to the surface of monolayers. The $\varepsilon_1$ defines the ability of a monolayer to store electric energy.
We find that the values for the static dielectric constant, $\varepsilon_1(0)$, (the values for the dielectric constant at zero energy) in E$_{\parallel}$ are $4.51$, $3.75$, $1.89$, and $1.7$; these values for E$_{\perp}$ are almost the same $1.49$, $1.47$, $1.36$, and $1.49$, for TlN, InN, GaN, and AlN monolayers, respectively. It seems that the InN monolayer has a better efficiency to store the electric energy at zero energy, when E$_{\rm \parallel}$ is considered, while the ability to store energy for all four considered monolayers is the same in the case of E$_{\rm \perp}$.

The value of $\varepsilon_1(0)$ and low value of $\varepsilon_1(\omega)$ are strongly
dependent on the band gaps of the monolayers. It has been shown that $\varepsilon_1(0)$ 
is inversely proportional to the band gap $\varepsilon_1(0) \approx 1/E_{g}$ \cite{PhysRev.128.2093}. 
We thus see that the $\varepsilon_1(0)$ value of TlN is maximum among all four considered monolayers as it's band gap is minimum and very small when E$_{\parallel}$ is considered. 
In addition, the refractive index, $n$, has almost the same qualitative characteristics as $\varepsilon_1$ in both directions for the electric field polarization.  
At zero energy, the value of $n(0)$ is directly proportional to $\varepsilon_1(0)$.
The value of $n(0)$ is equal to  $2.302$, $1.93$, $1.37$, and $1.306$  for TlN, GaN, InN and AlN, respectively, in the case of E$_{\parallel}$. Consequently, a maximum value of $n(0)$ for TlN is found. The $\varepsilon_1(0)$, and $n(0)$ agree very well with the literature \cite{VALEDBAGI2013153, ELAHI20169367}. 
\begin{figure}[htb]
	\centering
	\includegraphics[width=0.45\textwidth]{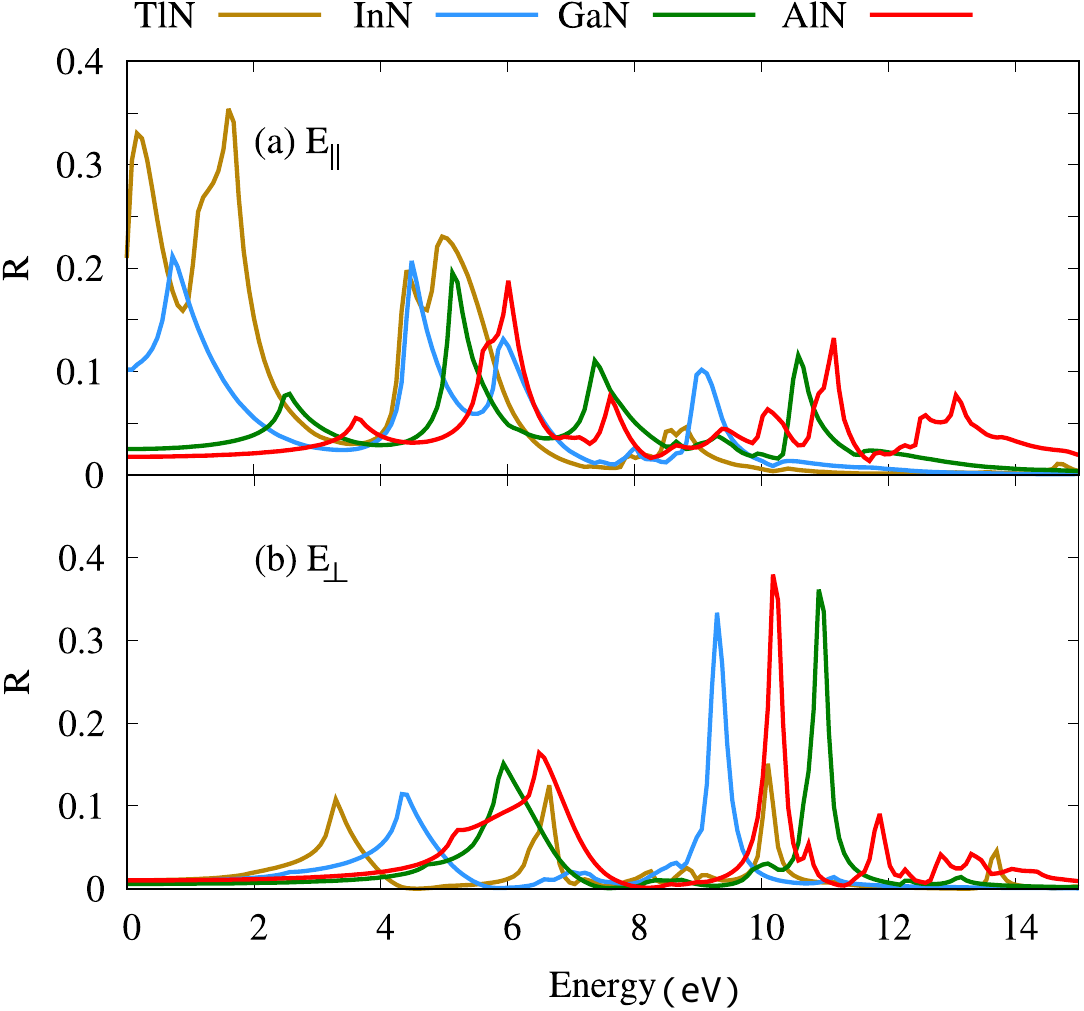}
	\caption{Reflectivity spectrum, $R$, for TlN (golden), InN (blue), GaN (green), and AlN (red) in the case of $E_{\rm \parallel}$ (a) and $E_{\rm \perp}$ (b).}
	\label{fig06}
\end{figure}

Another physical parameter of these monolayers is the optical reflectivity, $R(\omega)$, which introduces the ability of a monolayer to reflect light. The optical reflectivity of these monolayers for both E$_{\parallel}$ (a), and E$_{\perp}$ (b) in \fig{fig06}. We observe that the reflectivity of TlN and InN is high in the energy ranges $0\text{-}2.2$~eV when E$_{\parallel}$ is considered. This indicates that the transition is less (the monolayers are less transparent). In contrast, all four monolayers are almost $96\%$ transparent for the case of E$_{\perp}$ as their reflectivities are almost zero at this energy interval.
Furthermore, the reflectivity is decreased for TlN and InN in the visible light region, $1.65\text{-}3.1$~eV, while peaks in $R$ are seen for GaN and AlN for the same region of energy for the case of E$_{\parallel}$.

\begin{figure}[htb]
	\centering
	\includegraphics[width=0.45\textwidth]{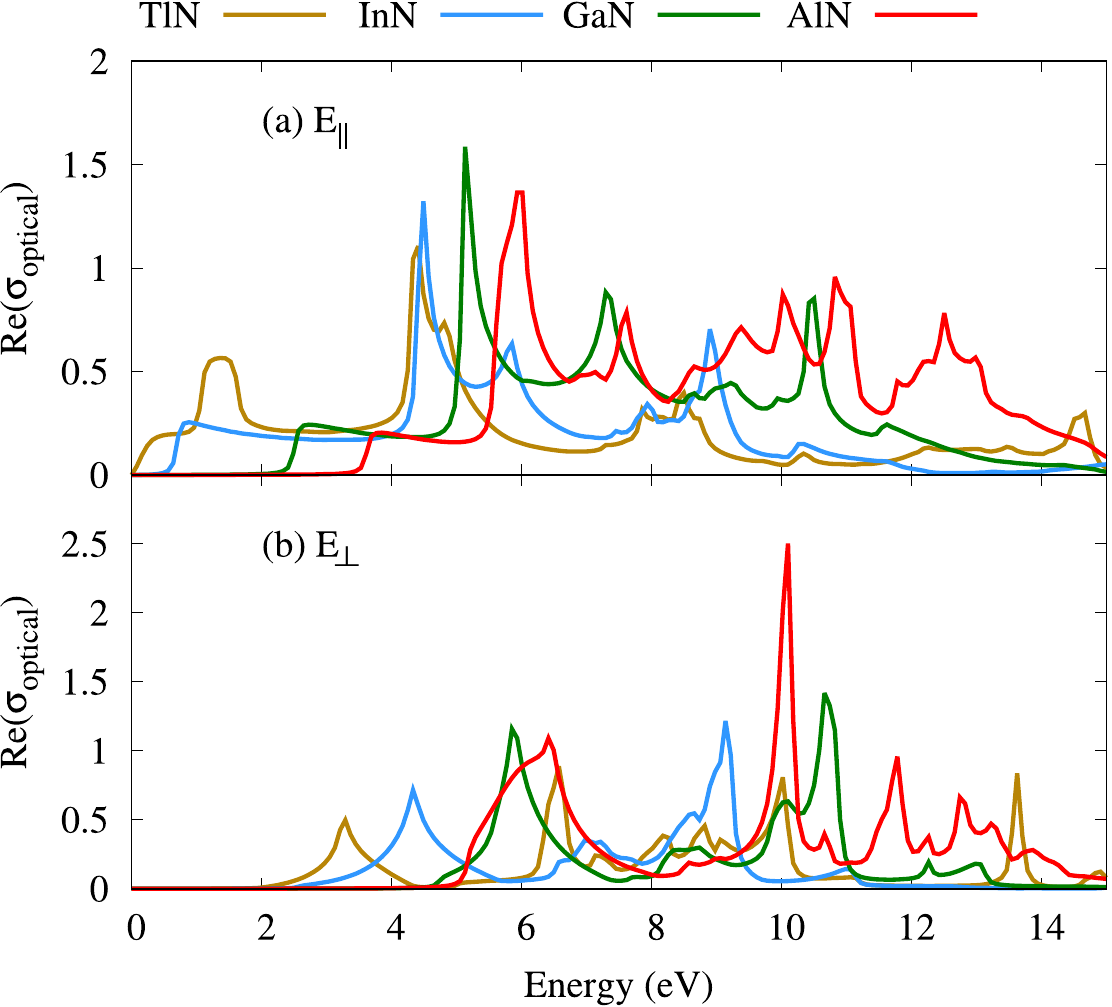}
	\caption{Real part of optical conductivity, $Re(\sigma_{\rm optical})$, for TlN (golden), InN (blue), GaN (green), and AlN (red) in the case of $E_{\rm \parallel}$ (a) and $E_{\rm \perp}$ (b).}
	\label{fig07}
\end{figure}

Finally, it is interesting to see the optical conductivity (real part) of these monolayers for both 
E$_{\parallel}$ (a), and E$_{\perp}$ (b) in \fig{fig07}. The optical gaps and the excitation energies of the monolayers can be clearly confirmed via the optical conductivity for both directions of polarization fotr the electric field. 

\section{Conclusion and remarks}\label{Sec:Conclusion}
The electronic and optical properties of metallic nitride compound monolayers were investigated using density functional theory. The optical properties of these monolayers were calculated using the random phase approximation implemented in the quantum espresso software. We have found both direct and indirect band gaps for the MN monolayers depending on the interatomic bond length between the M and the N atoms, and the contribution of the electronic orbitals of both the M and the N atoms. The optical properties calculated show that some of the MN monolayers have fascinating properties such as high transparency or reflectivity depending on the polarization of the incident light in the visible region, which could be very useful for optical nano-devices. 

\section{Acknowledgment}
This work was financially supported by the University of Sulaimani and 
the Research center of Komar University of Science and Technology. 
The computations were performed on resources provided by the Division of Computational 
Nanoscience at the University of Sulaimani.  
 


\begin{thebibliography}{10}
	\expandafter\ifx\csname url\endcsname\relax
	\def\url#1{\texttt{#1}}\fi
	\expandafter\ifx\csname urlprefix\endcsname\relax\def\urlprefix{URL }\fi
	\expandafter\ifx\csname href\endcsname\relax
	\def\href#1#2{#2} \def\path#1{#1}\fi
	
	\bibitem{Novoselov2005}
	K.~S. Novoselov, D.~Jiang, F.~Schedin, T.~J. Booth, V.~V. Khotkevich, S.~V.
	Morozov, A.~K. Geim,
	\href{http://www.pnas.org/content/102/30/10451.abstract}{Two-dimensional
		atomic crystals}, Proceedings of the National Academy of Sciences of the
	United States of America 102~(30) (2005) 10451.
	\newblock \href {https://doi.org/10.1073/pnas.0502848102}
	{\path{doi:10.1073/pnas.0502848102}}.
	\newline\urlprefix\url{http://www.pnas.org/content/102/30/10451.abstract}
	
	\bibitem{Li2021}
	W.~Li, X.~Qian, J.~Li, \href{https://doi.org/10.1038/s41578-021-00304-0}{Phase
		transitions in 2d materials}, Nature Reviews Materials 6~(9) (2021) 829--846.
	\newblock \href {https://doi.org/10.1038/s41578-021-00304-0}
	{\path{doi:10.1038/s41578-021-00304-0}}.
	\newline\urlprefix\url{https://doi.org/10.1038/s41578-021-00304-0}
	
	\bibitem{Wang2019}
	X.~Wang, Z.~Song, W.~Wen, H.~Liu, J.~Wu, C.~Dang, M.~Hossain, M.~A. Iqbal,
	L.~Xie,
	\href{https://onlinelibrary.wiley.com/doi/abs/10.1002/adma.201804682}{Potential
		2d materials with phase transitions: Structure, synthesis, and device
		applications}, Advanced Materials 31~(45) (2019) 1804682.
	\newblock \href
	{http://arxiv.org/abs/https://onlinelibrary.wiley.com/doi/pdf/10.1002/adma.201804682}
	{\path{arXiv:https://onlinelibrary.wiley.com/doi/pdf/10.1002/adma.201804682}},
	\href {https://doi.org/https://doi.org/10.1002/adma.201804682}
	{\path{doi:https://doi.org/10.1002/adma.201804682}}.
	\newline\urlprefix\url{https://onlinelibrary.wiley.com/doi/abs/10.1002/adma.201804682}
	
	\bibitem{RASHID2019102625}
	H.~O. Rashid, N.~R. Abdullah, V.~Gudmundsson,
	\href{http://www.sciencedirect.com/science/article/pii/S2211379719317140}{Silicon
		on a graphene nanosheet with triangle- and dot-shape: Electronic structure,
		specific heat, and thermal conductivity from first-principle calculations},
	Results in Physics 15 (2019) 102625.
	\newblock \href {https://doi.org/10.1016/j.rinp.2019.102625}
	{\path{doi:10.1016/j.rinp.2019.102625}}.
	\newline\urlprefix\url{http://www.sciencedirect.com/science/article/pii/S2211379719317140}
	
	\bibitem{Tan2020}
	T.~Tan, X.~Jiang, C.~Wang, B.~Yao, H.~Zhang,
	\href{https://onlinelibrary.wiley.com/doi/abs/10.1002/advs.202000058}{2d
		material optoelectronics for information functional device applications:
		Status and challenges}, Advanced Science 7~(11) (2020) 2000058.
	\newblock \href
	{http://arxiv.org/abs/https://onlinelibrary.wiley.com/doi/pdf/10.1002/advs.202000058}
	{\path{arXiv:https://onlinelibrary.wiley.com/doi/pdf/10.1002/advs.202000058}},
	\href {https://doi.org/https://doi.org/10.1002/advs.202000058}
	{\path{doi:https://doi.org/10.1002/advs.202000058}}.
	\newline\urlprefix\url{https://onlinelibrary.wiley.com/doi/abs/10.1002/advs.202000058}
	
	\bibitem{doi:10.1126/science.1102896}
	K.~S. Novoselov, A.~K. Geim, S.~V. Morozov, D.~Jiang, Y.~Zhang, S.~V. Dubonos,
	I.~V. Grigorieva, A.~A. Firsov,
	\href{https://www.science.org/doi/abs/10.1126/science.1102896}{Electric field
		effect in atomically thin carbon films}, Science 306~(5696) (2004) 666--669.
	\newblock \href
	{http://arxiv.org/abs/https://www.science.org/doi/pdf/10.1126/science.1102896}
	{\path{arXiv:https://www.science.org/doi/pdf/10.1126/science.1102896}}, \href
	{https://doi.org/10.1126/science.1102896}
	{\path{doi:10.1126/science.1102896}}.
	\newline\urlprefix\url{https://www.science.org/doi/abs/10.1126/science.1102896}
	
	\bibitem{ABDULLAH2020126350}
	N.~R. Abdullah, H.~O. Rashid, M.~T. Kareem, C.-S. Tang, A.~Manolescu,
	V.~Gudmundsson,
	\href{http://www.sciencedirect.com/science/article/pii/S0375960120301602}{Effects
		of bonded and non-bonded b/n codoping of graphene on its stability,
		interaction energy, electronic structure, and power factor}, Physics Letters
	A 384~(12) (2020) 126350.
	\newblock \href {https://doi.org/10.1016/j.physleta.2020.126350}
	{\path{doi:10.1016/j.physleta.2020.126350}}.
	\newline\urlprefix\url{http://www.sciencedirect.com/science/article/pii/S0375960120301602}
	
	\bibitem{ABDULLAH2020100740}
	N.~R. Abdullah, H.~O. Rashid, A.~Manolescu, V.~Gudmundsson,
	\href{http://www.sciencedirect.com/science/article/pii/S246802302030732X}{Interlayer
		interaction controlling the properties of ab- and aa-stacked bilayer
		graphene-like bc14n and si2c14}, Surfaces and Interfaces 21 (2020) 100740.
	\newblock \href {https://doi.org/https://doi.org/10.1016/j.surfin.2020.100740}
	{\path{doi:https://doi.org/10.1016/j.surfin.2020.100740}}.
	\newline\urlprefix\url{http://www.sciencedirect.com/science/article/pii/S246802302030732X}
	
	\bibitem{Balendhran2015}
	S.~Balendhran, S.~Walia, H.~Nili, S.~Sriram, M.~Bhaskaran,
	\href{https://onlinelibrary.wiley.com/doi/abs/10.1002/smll.201402041}{Elemental
		analogues of graphene: Silicene, germanene, stanene, and phosphorene}, Small
	11~(6) (2015) 640--652.
	\newblock \href
	{http://arxiv.org/abs/https://onlinelibrary.wiley.com/doi/pdf/10.1002/smll.201402041}
	{\path{arXiv:https://onlinelibrary.wiley.com/doi/pdf/10.1002/smll.201402041}},
	\href {https://doi.org/https://doi.org/10.1002/smll.201402041}
	{\path{doi:https://doi.org/10.1002/smll.201402041}}.
	\newline\urlprefix\url{https://onlinelibrary.wiley.com/doi/abs/10.1002/smll.201402041}
	
	\bibitem{Xu2013}
	M.~Xu, T.~Liang, M.~Shi, H.~Chen,
	\href{https://doi.org/10.1021/cr300263a}{Graphene-like two-dimensional
		materials}, Chemical Reviews 113~(5) (2013) 3766--3798.
	\newblock \href {https://doi.org/10.1021/cr300263a}
	{\path{doi:10.1021/cr300263a}}.
	\newline\urlprefix\url{https://doi.org/10.1021/cr300263a}
	
	\bibitem{doi:10.1021/acs.chemrev.6b00558}
	C.~Tan, X.~Cao, X.-J. Wu, Q.~He, J.~Yang, X.~Zhang, J.~Chen, W.~Zhao, S.~Han,
	G.-H. Nam, M.~Sindoro, H.~Zhang,
	\href{https://doi.org/10.1021/acs.chemrev.6b00558}{Recent advances in
		ultrathin two-dimensional nanomaterials}, Chemical Reviews 117~(9) (2017)
	6225--6331, pMID: 28306244.
	\newblock \href
	{http://arxiv.org/abs/https://doi.org/10.1021/acs.chemrev.6b00558}
	{\path{arXiv:https://doi.org/10.1021/acs.chemrev.6b00558}}, \href
	{https://doi.org/10.1021/acs.chemrev.6b00558}
	{\path{doi:10.1021/acs.chemrev.6b00558}}.
	\newline\urlprefix\url{https://doi.org/10.1021/acs.chemrev.6b00558}
	
	\bibitem{Prete2020}
	M.~S. Prete, D.~Grassano, O.~Pulci, I.~Kupchak, V.~Olevano, F.~Bechstedt,
	\href{https://doi.org/10.1038/s41598-020-67667-2}{Giant excitonic absorption
		and emission in two-dimensional group-iii nitrides}, Scientific Reports
	10~(1) (2020) 10719.
	\newblock \href {https://doi.org/10.1038/s41598-020-67667-2}
	{\path{doi:10.1038/s41598-020-67667-2}}.
	\newline\urlprefix\url{https://doi.org/10.1038/s41598-020-67667-2}
	
	\bibitem{doi:10.1063/1.3041639}
	W.-Q. Han, L.~Wu, Y.~Zhu, K.~Watanabe, T.~Taniguchi,
	\href{https://doi.org/10.1063/1.3041639}{Structure of chemically derived
		mono- and few-atomic-layer boron nitride sheets}, Applied Physics Letters
	93~(22) (2008) 223103.
	\newblock \href {http://arxiv.org/abs/https://doi.org/10.1063/1.3041639}
	{\path{arXiv:https://doi.org/10.1063/1.3041639}}, \href
	{https://doi.org/10.1063/1.3041639} {\path{doi:10.1063/1.3041639}}.
	\newline\urlprefix\url{https://doi.org/10.1063/1.3041639}
	
	\bibitem{pub.1014753295}
	Y.~Stehle, H.~M. Meyer, R.~R. Unocic, M.~Kidder, G.~Polizos, P.~G. Datskos,
	R.~Jackson, S.~N. Smirnov, I.~V. Vlassiouk,
	\href{https://app.dimensions.ai/details/publication/pub.1014753295}{Synthesis
		of hexagonal boron nitride monolayer: Control of nucleation and crystal
		morphology}, Chemistry of Materials 27~(23) (2015) 8041--8047.
	\newblock \href {https://doi.org/10.1021/acs.chemmater.5b03607}
	{\path{doi:10.1021/acs.chemmater.5b03607}}.
	\newline\urlprefix\url{https://app.dimensions.ai/details/publication/pub.1014753295}
	
	\bibitem{PhysRevB.79.115442}
	M.~Topsakal, E.~Akt\"urk, S.~Ciraci,
	\href{https://link.aps.org/doi/10.1103/PhysRevB.79.115442}{First-principles
		study of two- and one-dimensional honeycomb structures of boron nitride},
	Phys. Rev. B 79 (2009) 115442.
	\newblock \href {https://doi.org/10.1103/PhysRevB.79.115442}
	{\path{doi:10.1103/PhysRevB.79.115442}}.
	\newline\urlprefix\url{https://link.aps.org/doi/10.1103/PhysRevB.79.115442}
	
	\bibitem{C2CP40081B}
	Y.~Zhao, X.~Wu, J.~Yang, X.~C. Zeng,
	\href{http://dx.doi.org/10.1039/C2CP40081B}{Oxidation of a two-dimensional
		hexagonal boron nitride monolayer: a first-principles study}, Phys. Chem.
	Chem. Phys. 14 (2012) 5545--5550.
	\newblock \href {https://doi.org/10.1039/C2CP40081B}
	{\path{doi:10.1039/C2CP40081B}}.
	\newline\urlprefix\url{http://dx.doi.org/10.1039/C2CP40081B}
	
	\bibitem{C2NR32366D}
	Q.~Peng, W.~Ji, S.~De,
	\href{http://dx.doi.org/10.1039/C2NR32366D}{First-principles study of the
		effects of mechanical strains on the radiation hardness of hexagonal boron
		nitride monolayers}, Nanoscale 5 (2013) 695--703.
	\newblock \href {https://doi.org/10.1039/C2NR32366D}
	{\path{doi:10.1039/C2NR32366D}}.
	\newline\urlprefix\url{http://dx.doi.org/10.1039/C2NR32366D}
	
	\bibitem{ZHANG2007317}
	X.~Zhang, Z.~Liu, S.~Hark,
	\href{https://www.sciencedirect.com/science/article/pii/S0038109807004322}{Synthesis
		and optical characterization of single-crystalline aln nanosheets}, Solid
	State Communications 143~(6) (2007) 317--320.
	\newblock \href {https://doi.org/https://doi.org/10.1016/j.ssc.2007.05.039}
	{\path{doi:https://doi.org/10.1016/j.ssc.2007.05.039}}.
	\newline\urlprefix\url{https://www.sciencedirect.com/science/article/pii/S0038109807004322}
	
	\bibitem{doi:10.1063/1.4851239}
	P.~Tsipas, S.~Kassavetis, D.~Tsoutsou, E.~Xenogiannopoulou, E.~Golias, S.~A.
	Giamini, C.~Grazianetti, D.~Chiappe, A.~Molle, M.~Fanciulli, A.~Dimoulas,
	\href{https://doi.org/10.1063/1.4851239}{Evidence for graphite-like hexagonal
		aln nanosheets epitaxially grown on single crystal ag(111)}, Applied Physics
	Letters 103~(25) (2013) 251605.
	\newblock \href {http://arxiv.org/abs/https://doi.org/10.1063/1.4851239}
	{\path{arXiv:https://doi.org/10.1063/1.4851239}}, \href
	{https://doi.org/10.1063/1.4851239} {\path{doi:10.1063/1.4851239}}.
	\newline\urlprefix\url{https://doi.org/10.1063/1.4851239}
	
	\bibitem{Mansurov2015}
	V.~Mansurov, T.~Malin, Y.~Galitsyn, K.~Zhuravlev, Graphene-like aln layer
	formation on (111)si surface by ammonia molecular beam epitaxy, Journal of
	Crystal Growth 428~(C) (2015) 93--97.
	\newblock \href {https://doi.org/10.1016/j.jcrysgro.2015.07.030}
	{\path{doi:10.1016/j.jcrysgro.2015.07.030}}.
	
	\bibitem{Seo2015}
	T.~H. Seo, A.~H. Park, S.~Park, Y.~H. Kim, G.~H. Lee, M.~J. Kim, M.~S. Jeong,
	Y.~H. Lee, Y.-B. Hahn, E.-K. Suh,
	\href{https://doi.org/10.1038/srep07747}{Direct growth of gan layer on carbon
		nanotube-graphene hybrid structure and its application for light emitting
		diodes}, Scientific Reports 5~(1) (2015) 7747.
	\newblock \href {https://doi.org/10.1038/srep07747}
	{\path{doi:10.1038/srep07747}}.
	\newline\urlprefix\url{https://doi.org/10.1038/srep07747}
	
	\bibitem{AlBalushi2016}
	Z.~Y. Al~Balushi, K.~Wang, R.~K. Ghosh, R.~A. Vil{\'a}, S.~M. Eichfeld, J.~D.
	Caldwell, X.~Qin, Y.-C. Lin, P.~A. DeSario, G.~Stone, S.~Subramanian, D.~F.
	Paul, R.~M. Wallace, S.~Datta, J.~Redwing, J.~A. Robinson,
	\href{https://doi.org/10.1038/nmat4742}{Two-dimensional gallium nitride
		realized via graphene encapsulation}, Nature Materials 15~(11) (2016)
	1166--1171.
	\newblock \href {https://doi.org/10.1038/nmat4742}
	{\path{doi:10.1038/nmat4742}}.
	\newline\urlprefix\url{https://doi.org/10.1038/nmat4742}
	
	\bibitem{doi:10.1063/1.2712801}
	H.~Y. Xu, Z.~Liu, X.~T. Zhang, S.~K. Hark,
	\href{https://doi.org/10.1063/1.2712801}{Synthesis and optical properties of
		inn nanowires and nanotubes}, Applied Physics Letters 90~(11) (2007) 113105.
	\newblock \href {http://arxiv.org/abs/https://doi.org/10.1063/1.2712801}
	{\path{arXiv:https://doi.org/10.1063/1.2712801}}, \href
	{https://doi.org/10.1063/1.2712801} {\path{doi:10.1063/1.2712801}}.
	\newline\urlprefix\url{https://doi.org/10.1063/1.2712801}
	
	\bibitem{doi:10.1021/nl4030819}
	S.~Zhao, B.~H. Le, D.~P. Liu, X.~D. Liu, M.~G. Kibria, T.~Szkopek, H.~Guo,
	Z.~Mi, \href{https://doi.org/10.1021/nl4030819}{p-type inn nanowires}, Nano
	Letters 13~(11) (2013) 5509--5513, pMID: 24090401.
	\newblock \href {http://arxiv.org/abs/https://doi.org/10.1021/nl4030819}
	{\path{arXiv:https://doi.org/10.1021/nl4030819}}, \href
	{https://doi.org/10.1021/nl4030819} {\path{doi:10.1021/nl4030819}}.
	\newline\urlprefix\url{https://doi.org/10.1021/nl4030819}
	
	\bibitem{doi:10.1063/1.4967928}
	A.~Yoshikawa, K.~Kusakabe, N.~Hashimoto, E.-S. Hwang, D.~Imai, T.~Itoi,
	\href{https://doi.org/10.1063/1.4967928}{Systematic study on dynamic atomic
		layer epitaxy of inn on/in +c-gan matrix and fabrication of fine-structure
		inn/gan quantum wells: Role of high growth temperature}, Journal of Applied
	Physics 120~(22) (2016) 225303.
	\newblock \href {http://arxiv.org/abs/https://doi.org/10.1063/1.4967928}
	{\path{arXiv:https://doi.org/10.1063/1.4967928}}, \href
	{https://doi.org/10.1063/1.4967928} {\path{doi:10.1063/1.4967928}}.
	\newline\urlprefix\url{https://doi.org/10.1063/1.4967928}
	
	\bibitem{SHI2010203}
	L.~Shi, Y.~Duan, L.~Qin,
	\href{https://www.sciencedirect.com/science/article/pii/S0927025610004532}{Structural
		phase transition, electronic and elastic properties in tlx (x=n, p, as)
		compounds: Pressure-induced effects}, Computational Materials Science 50~(1)
	(2010) 203--210.
	\newblock \href
	{https://doi.org/https://doi.org/10.1016/j.commatsci.2010.07.027}
	{\path{doi:https://doi.org/10.1016/j.commatsci.2010.07.027}}.
	\newline\urlprefix\url{https://www.sciencedirect.com/science/article/pii/S0927025610004532}
	
	\bibitem{Elahi2016}
	S.~Elahi, M.~Farzan, H.~Salehi, M.~Abolhasani, Original research article, Optik
	- International Journal for Light and Electron Optics 127~(20) (2016)
	9367--9376.
	\newblock \href {https://doi.org/10.1016/j.ijleo.2016.07.013}
	{\path{doi:10.1016/j.ijleo.2016.07.013}}.
	
	\bibitem{Wei2010}
	S.~Li-Wei, D.~Yi-Feng, Q.~Li-Xia,
	\href{https://doi.org/10.1088/0256-307x/27/8/080505}{Structural stability and
		elastic properties of wurtzite {TIN} under hydrostatic pressure} 27~(8)
	(2010) 080505.
	\newblock \href {https://doi.org/10.1088/0256-307x/27/8/080505}
	{\path{doi:10.1088/0256-307x/27/8/080505}}.
	\newline\urlprefix\url{https://doi.org/10.1088/0256-307x/27/8/080505}
	
	\bibitem{VALEDBAGI2013153}
	S.~Valedbagi, A.~Fathalian, S.~{Mohammad Elahi},
	\href{https://www.sciencedirect.com/science/article/pii/S0030401813006251}{Electronic
		and optical properties of aln nanosheet: An ab initio study}, Optics
	Communications 309 (2013) 153--157.
	\newblock \href {https://doi.org/https://doi.org/10.1016/j.optcom.2013.06.061}
	{\path{doi:https://doi.org/10.1016/j.optcom.2013.06.061}}.
	\newline\urlprefix\url{https://www.sciencedirect.com/science/article/pii/S0030401813006251}
	
	\bibitem{VAHEDIFAKHRABAD201538}
	D.~{Vahedi Fakhrabad}, N.~Shahtahmasebi, M.~Ashhadi,
	\href{https://www.sciencedirect.com/science/article/pii/S0749603614004807}{Optical
		excitations and quasiparticle energies in the aln monolayer honeycomb
		structure}, Superlattices and Microstructures 79 (2015) 38--44.
	\newblock \href {https://doi.org/https://doi.org/10.1016/j.spmi.2014.12.012}
	{\path{doi:https://doi.org/10.1016/j.spmi.2014.12.012}}.
	\newline\urlprefix\url{https://www.sciencedirect.com/science/article/pii/S0749603614004807}
	
	\bibitem{Meng2016}
	R.~Meng, J.~Jiang, Q.~Liang, Q.~Yang, C.~Tan, X.~Sun, X.~Chen,
	\href{https://doi.org/10.1007/s40843-016-5122-3}{Design of graphene-like
		gallium nitride and ws2/wse2 nanocomposites for photocatalyst applications},
	Science China Materials 59~(12) (2016) 1027--1036.
	\newblock \href {https://doi.org/10.1007/s40843-016-5122-3}
	{\path{doi:10.1007/s40843-016-5122-3}}.
	\newline\urlprefix\url{https://doi.org/10.1007/s40843-016-5122-3}
	
	\bibitem{C8CP05529G}
	A.~Mogulkoc, Y.~Mogulkoc, M.~Modarresi, B.~Alkan,
	\href{http://dx.doi.org/10.1039/C8CP05529G}{Electronic structure and optical
		properties of novel monolayer gallium nitride and boron phosphide
		heterobilayers}, Phys. Chem. Chem. Phys. 20 (2018) 28124--28134.
	\newblock \href {https://doi.org/10.1039/C8CP05529G}
	{\path{doi:10.1039/C8CP05529G}}.
	\newline\urlprefix\url{http://dx.doi.org/10.1039/C8CP05529G}
	
	\bibitem{DU2021110008}
	K.~Du, Z.~Xiong, L.~Ao, L.~Chen,
	\href{https://www.sciencedirect.com/science/article/pii/S0042207X20308678}{Tuning
		the electronic and optical properties of two-dimensional gallium nitride by
		chemical functionalization}, Vacuum 185 (2021) 110008.
	\newblock \href {https://doi.org/https://doi.org/10.1016/j.vacuum.2020.110008}
	{\path{doi:https://doi.org/10.1016/j.vacuum.2020.110008}}.
	\newline\urlprefix\url{https://www.sciencedirect.com/science/article/pii/S0042207X20308678}
	
	\bibitem{D0RA01025A}
	T.~V. Vu, K.~D. Pham, T.~N. Pham, D.~D. Vo, P.~T. Dang, C.~V. Nguyen, H.~V.
	Phuc, N.~T.~T. Binh, D.~M. Hoat, N.~N. Hieu,
	\href{http://dx.doi.org/10.1039/D0RA01025A}{First-principles prediction of
		chemically functionalized inn monolayers: electronic and optical properties},
	RSC Adv. 10 (2020) 10731--10739.
	\newblock \href {https://doi.org/10.1039/D0RA01025A}
	{\path{doi:10.1039/D0RA01025A}}.
	\newline\urlprefix\url{http://dx.doi.org/10.1039/D0RA01025A}
	
	\bibitem{FERREIRADASILVA2005151}
	A.~{Ferreira da Silva}, N.~{Souza Dantas}, J.~{de Almeida}, R.~Ahuja,
	C.~Persson,
	\href{https://www.sciencedirect.com/science/article/pii/S0022024805003167}{Electronic
		and optical properties of wurtzite and zinc-blende tln and aln}, Journal of
	Crystal Growth 281~(1) (2005) 151--160, the Internbational Workshop on Bulk
	Nitride Semiconductors III.
	\newblock \href
	{https://doi.org/https://doi.org/10.1016/j.jcrysgro.2005.03.021}
	{\path{doi:https://doi.org/10.1016/j.jcrysgro.2005.03.021}}.
	\newline\urlprefix\url{https://www.sciencedirect.com/science/article/pii/S0022024805003167}
	
	\bibitem{ABDULLAH2021110095}
	N.~R. Abdullah, H.~O. Rashid, C.-S. Tang, A.~Manolescu, V.~Gudmundsson,
	\href{https://www.sciencedirect.com/science/article/pii/S002236972100161X}{Role
		of interlayer spacing on electronic, thermal and optical properties of
		bn-codoped bilayer graphene: Influence of the interlayer and the induced
		dipole-dipole interactions}, Journal of Physics and Chemistry of Solids 155
	(2021) 110095.
	\newblock \href {https://doi.org/https://doi.org/10.1016/j.jpcs.2021.110095}
	{\path{doi:https://doi.org/10.1016/j.jpcs.2021.110095}}.
	\newline\urlprefix\url{https://www.sciencedirect.com/science/article/pii/S002236972100161X}
	
	\bibitem{Giannozzi_2009}
	P.~Giannozzi, S.~Baroni, N.~Bonini, M.~Calandra, R.~Car, C.~Cavazzoni,
	D.~Ceresoli, G.~L. Chiarotti, M.~Cococcioni, I.~Dabo, A.~D. Corso,
	S.~de~Gironcoli, S.~Fabris, G.~Fratesi, R.~Gebauer, U.~Gerstmann,
	C.~Gougoussis, A.~Kokalj, M.~Lazzeri, L.~Martin-Samos, N.~Marzari, F.~Mauri,
	R.~Mazzarello, S.~Paolini, A.~Pasquarello, L.~Paulatto, C.~Sbraccia,
	S.~Scandolo, G.~Sclauzero, A.~P. Seitsonen, A.~Smogunov, P.~Umari, R.~M.
	Wentzcovitch,
	\href{https://doi.org/10.1088%2F0953-8984%2F21%2F39%2F395502}{{QUANTUM}
		{ESPRESSO}: a modular and open-source software project for quantum
		simulations of materials}, Journal of Physics: Condensed Matter 21~(39)
	(2009) 395502.
	\newblock \href {https://doi.org/10.1088/0953-8984/21/39/395502}
	{\path{doi:10.1088/0953-8984/21/39/395502}}.
	\newline\urlprefix\url{https://doi.org/10.1088%2F0953-8984%2F21%2F39%2F395502}
	
	\bibitem{giannozzi2017advanced}
	P.~Giannozzi, O.~Andreussi, T.~Brumme, O.~Bunau, M.~B. Nardelli, M.~Calandra,
	R.~Car, C.~Cavazzoni, D.~Ceresoli, M.~Cococcioni, et~al., Advanced
	capabilities for materials modelling with quantum espresso, Journal of
	Physics: Condensed Matter 29~(46) (2017) 465901.
	
	\bibitem{ABDULLAH2020126807}
	N.~R. Abdullah, H.~O. Rashid, C.-S. Tang, A.~Manolescu, V.~Gudmundsson,
	\href{http://www.sciencedirect.com/science/article/pii/S0375960120306745}{Modeling
		electronic, mechanical, optical and thermal properties of graphene-like bc6n
		materials: Role of prominent bn-bonds}, Physics Letters A 384~(32) (2020)
	126807.
	\newblock \href
	{https://doi.org/https://doi.org/10.1016/j.physleta.2020.126807}
	{\path{doi:https://doi.org/10.1016/j.physleta.2020.126807}}.
	\newline\urlprefix\url{http://www.sciencedirect.com/science/article/pii/S0375960120306745}
	
	\bibitem{ABDULLAH2020103282}
	N.~R. Abdullah, D.~A. Abdalla, T.~Y. Ahmed, S.~W. Abdulqadr, H.~O. Rashid,
	\href{http://www.sciencedirect.com/science/article/pii/S2211379720317496}{Effect
		of bn dimers on the stability, electronic, and thermal properties of
		monolayer graphene}, Results in Physics 18 (2020) 103282.
	\newblock \href {https://doi.org/https://doi.org/10.1016/j.rinp.2020.103282}
	{\path{doi:https://doi.org/10.1016/j.rinp.2020.103282}}.
	\newline\urlprefix\url{http://www.sciencedirect.com/science/article/pii/S2211379720317496}
	
	\bibitem{ELAHI20169367}
	S.~Elahi, M.~Farzan, H.~Salehi, M.~Abolhasani,
	\href{https://www.sciencedirect.com/science/article/pii/S0030402616307781}{An
		investigation of electronic and optical properties of tln nanosheet and
		compare with tln bulk (wurtzite) by first principle}, Optik 127~(20) (2016)
	9367--9376.
	\newblock \href {https://doi.org/https://doi.org/10.1016/j.ijleo.2016.07.013}
	{\path{doi:https://doi.org/10.1016/j.ijleo.2016.07.013}}.
	\newline\urlprefix\url{https://www.sciencedirect.com/science/article/pii/S0030402616307781}
	
	\bibitem{ABDULLAH2020114556}
	N.~R. Abdullah, H.~O. Rashid, C.-S. Tang, A.~Manolescu, V.~Gudmundsson,
	\href{http://www.sciencedirect.com/science/article/pii/S1386947720316246}{Properties
		of bsi6n monolayers derived by first-principle computation}, Physica E:
	Low-dimensional Systems and Nanostructures (2020) 114556\href
	{https://doi.org/https://doi.org/10.1016/j.physe.2020.114556}
	{\path{doi:https://doi.org/10.1016/j.physe.2020.114556}}.
	\newline\urlprefix\url{http://www.sciencedirect.com/science/article/pii/S1386947720316246}
	
	\bibitem{ABDULLAH2021114644}
	N.~R. Abdullah, M.~T. Kareem, H.~O. Rashid, A.~Manolescu, V.~Gudmundsson,
	\href{https://www.sciencedirect.com/science/article/pii/S1386947721000266}{Spin-polarised
		dft modeling of electronic, magnetic, thermal and optical properties of
		silicene doped with transition metals}, Physica E: Low-dimensional Systems
	and Nanostructures 129 (2021) 114644.
	\newblock \href {https://doi.org/https://doi.org/10.1016/j.physe.2021.114644}
	{\path{doi:https://doi.org/10.1016/j.physe.2021.114644}}.
	\newline\urlprefix\url{https://www.sciencedirect.com/science/article/pii/S1386947721000266}
	
	\bibitem{Liu2019}
	X.-F. Liu, Z.-J. Luo, X.~Zhou, J.-M. Wei, Y.~Wang, X.~Guo, B.~Lv, Z.~Ding,
	\href{https://doi.org/10.1088/1674-1056/28/8/086105}{Structural, mechanical,
		and electronic properties of 25 kinds of {III}{\textendash}v binary
		monolayers: A computational study with first-principles calculation} 28~(8)
	(2019) 086105.
	\newblock \href {https://doi.org/10.1088/1674-1056/28/8/086105}
	{\path{doi:10.1088/1674-1056/28/8/086105}}.
	\newline\urlprefix\url{https://doi.org/10.1088/1674-1056/28/8/086105}
	
	\bibitem{ABDULLAH2021106073}
	N.~R. Abdullah, B.~J. Abdullah, C.-S. Tang, V.~Gudmundsson,
	\href{https://www.sciencedirect.com/science/article/pii/S1369800121004182}{Properties
		of bc6n monolayer derived by first-principle computation: Influences of
		interactions between dopant atoms on thermoelectric and optical properties},
	Materials Science in Semiconductor Processing 135 (2021) 106073.
	\newblock \href {https://doi.org/https://doi.org/10.1016/j.mssp.2021.106073}
	{\path{doi:https://doi.org/10.1016/j.mssp.2021.106073}}.
	\newline\urlprefix\url{https://www.sciencedirect.com/science/article/pii/S1369800121004182}
	
	\bibitem{doi:10.1063/1.1564060}
	J.~Heyd, G.~E. Scuseria, M.~Ernzerhof,
	\href{https://doi.org/10.1063/1.1564060}{Hybrid functionals based on a
		screened coulomb potential}, The Journal of Chemical Physics 118~(18) (2003)
	8207--8215.
	\newblock \href {http://arxiv.org/abs/https://doi.org/10.1063/1.1564060}
	{\path{arXiv:https://doi.org/10.1063/1.1564060}}, \href
	{https://doi.org/10.1063/1.1564060} {\path{doi:10.1063/1.1564060}}.
	\newline\urlprefix\url{https://doi.org/10.1063/1.1564060}
	
	\bibitem{PhysRev.139.A796}
	L.~Hedin, \href{https://link.aps.org/doi/10.1103/PhysRev.139.A796}{New method
		for calculating the one-particle green's function with application to the
		electron-gas problem}, Phys. Rev. 139 (1965) A796--A823.
	\newblock \href {https://doi.org/10.1103/PhysRev.139.A796}
	{\path{doi:10.1103/PhysRev.139.A796}}.
	\newline\urlprefix\url{https://link.aps.org/doi/10.1103/PhysRev.139.A796}
	
	\bibitem{PhysRevB.50.4397}
	N.~E. Christensen, I.~Gorczyca,
	\href{https://link.aps.org/doi/10.1103/PhysRevB.50.4397}{Optical and
		structural properties of iii-v nitrides under pressure}, Phys. Rev. B 50
	(1994) 4397--4415.
	\newblock \href {https://doi.org/10.1103/PhysRevB.50.4397}
	{\path{doi:10.1103/PhysRevB.50.4397}}.
	\newline\urlprefix\url{https://link.aps.org/doi/10.1103/PhysRevB.50.4397}
	
	\bibitem{JOHN2017307}
	R.~John, B.~Merlin,
	\href{http://www.sciencedirect.com/science/article/pii/S0022369717300367}{Optical
		properties of graphene, silicene, germanene, and stanene from ir to far uv
		– a first principles study}, Journal of Physics and Chemistry of Solids 110
	(2017) 307 -- 315.
	\newblock \href {https://doi.org/https://doi.org/10.1016/j.jpcs.2017.06.026}
	{\path{doi:https://doi.org/10.1016/j.jpcs.2017.06.026}}.
	\newline\urlprefix\url{http://www.sciencedirect.com/science/article/pii/S0022369717300367}
	
	\bibitem{Ren2012}
	X.~Ren, P.~Rinke, C.~Joas, M.~Scheffler,
	\href{https://doi.org/10.1007/s10853-012-6570-4}{Random-phase approximation
		and its applications in computational chemistry and materials science},
	Journal of Materials Science 47~(21) (2012) 7447--7471.
	\newblock \href {https://doi.org/10.1007/s10853-012-6570-4}
	{\path{doi:10.1007/s10853-012-6570-4}}.
	\newline\urlprefix\url{https://doi.org/10.1007/s10853-012-6570-4}
	
	\bibitem{ABDULLAH2020126578}
	N.~R. Abdullah, G.~A. Mohammed, H.~O. Rashid, V.~Gudmundsson,
	\href{http://www.sciencedirect.com/science/article/pii/S037596012030445X}{Electronic,
		thermal, and optical properties of graphene like sicx structures: Significant
		effects of si atom configurations}, Physics Letters A 384~(24) (2020) 126578.
	\newblock \href
	{https://doi.org/https://doi.org/10.1016/j.physleta.2020.126578}
	{\path{doi:https://doi.org/10.1016/j.physleta.2020.126578}}.
	\newline\urlprefix\url{http://www.sciencedirect.com/science/article/pii/S037596012030445X}
	
	\bibitem{PhysRev.128.2093}
	D.~R. Penn,
	\href{https://link.aps.org/doi/10.1103/PhysRev.128.2093}{Wave-number-dependent
		dielectric function of semiconductors}, Phys. Rev. 128 (1962) 2093--2097.
	\newblock \href {https://doi.org/10.1103/PhysRev.128.2093}
	{\path{doi:10.1103/PhysRev.128.2093}}.
	\newline\urlprefix\url{https://link.aps.org/doi/10.1103/PhysRev.128.2093}
	
\end{thebibliography}

\end{document}